%% file: sample-sigconf.tex
\documentclass[sigconf]{acmart}

\AtBeginDocument{%
  }

\setcopyright{acmlicensed}
\copyrightyear{2025}
\acmYear{2025}
\acmDOI{10.1145/3701551.3703552}

\acmConference[WSDM '25] {Proceedings of the Eighteenth ACM International Conference on Web Search and Data Mining}{March 10--14, 2025}{Hannover, Germany.}
\acmBooktitle{Proceedings of the Eighteenth ACM International Conference on Web Search and Data Mining (WSDM '25), March 10--14, 2025, Hannover, Germany}
\acmISBN{979-8-4007-1329-3/25/03}

\newcommand{\zhang}{}
\newcommand{\cheng}{}

\usepackage{multirow}
\usepackage[normalem]{ulem}
\usepackage{graphicx}
\usepackage{subfig}
\usepackage{enumitem}
\useunder{\uline}{\ul}{}
\begin{document}

\title{Facet-Aware Multi-Head Mixture-of-Experts Model for Sequential Recommendation}


\author{Mingrui Liu}
\affiliation{%
  \institution{Nanyang Technological University}
  \country{Singapore}}
\email{mingrui001@e.ntu.edu.sg}

\author{Sixiao Zhang}
\authornote{Co-corresponding authors.}
\affiliation{%
  \institution{Nanyang Technological University}
  \country{Singapore}}
\email{sixiao001@e.ntu.edu.sg}

\author{Cheng Long}
\authornotemark[1]
\affiliation{%
  \institution{Nanyang Technological University}
  \country{Singapore}}
\email{c.long@ntu.edu.sg}

\renewcommand{\shortauthors}{Mingrui Liu, Sixiao Zhang, and Cheng Long}

\begin{abstract}
  \input{abstract}
\end{abstract}




\maketitle

\input{introduction}

\input{relatedwork}

\input{Preliminaries}

\input{Methods}

\input{experiments}

\input{conclusion}

\begin{acks}
This research is supported by the Ministry of Education, Singapore, under its Academic Research Fund (Tier 2 Award MOE-T2EP20221-0013, Tier 2 Award MOE-T2EP20220-0011, and Tier 1 Award (RG20/24)). Any opinions, findings and conclusions or recommendations expressed in this material are those of the author(s) and do not reflect the views of the Ministry of Education, Singapore.
\end{acks}

\bibliographystyle{ACM-Reference-Format}
\bibliography{sample-base}


\end{document}

%% file: abstract.tex
Sequential recommendation (SR) systems excel at capturing users' dynamic preferences by leveraging their interaction histories.
Most existing SR systems assign a single embedding vector to each item to represent its features, and various types of models are adopted to combine these item embeddings into a sequence representation vector to capture the user intent. 
However, we argue that this representation {\zhang alone} is insufficient to capture an item's multi-faceted nature (e.g., movie genres, starring actors). Besides, users often exhibit complex and varied preferences within these facets (e.g., liking both action and musical films in the facet of genre), which are challenging to fully represent.
To address the issues above, we propose a novel structure called
\textit{\textbf{\underline{F}}acet-\textbf{\underline{A}}ware \textbf{\underline{M}}ulti-Head Mixture-of-\textbf{\underline{E}}xperts Model for Sequential Recommendation} (\textbf{\textit{FAME}}).
We leverage sub-embeddings from each head in the last multi-head attention layer to predict the next item separately. A gating mechanism integrates recommendations from each head {\zhang and} dynamically {\zhang determines} their importance.
Furthermore, we introduce a Mixture-of-Experts (MoE) network in each attention head to disentangle various user preferences within each facet. Each expert within the MoE focuses on a specific preference. {\zhang A learnable router network is adopted to compute the importance weight for each expert and aggregate them.} 
We conduct extensive experiments on four public sequential recommendation datasets and the
results demonstrate the effectiveness of our method over existing baseline models. 

%% file: introduction.tex
\section{Introduction}
The explosion of information online presents users with a vast and ever-growing sea of items, from products~\cite{product} and apps~\cite{content} to videos~\cite{video1, video2}. With limited time to explore everything, recommender systems (RS) have become crucial tools for helping users make efficient and satisfying choices. However, user interests are inherently dynamic, evolving over time and making it challenging for platforms to deliver consistently relevant recommendations~\cite{wang2021survey}. 
To address these challenges, sequential recommendation (SR) has emerged as a powerful technique. This approach leverages the sequential nature of user interactions, typically captured as sessions containing a series of recent item interactions, to predict the user's next action~\cite{fang2020deep, wang2019sequential}.

\begin{figure}[htbp]
  \centering
  \includegraphics[width=\linewidth]{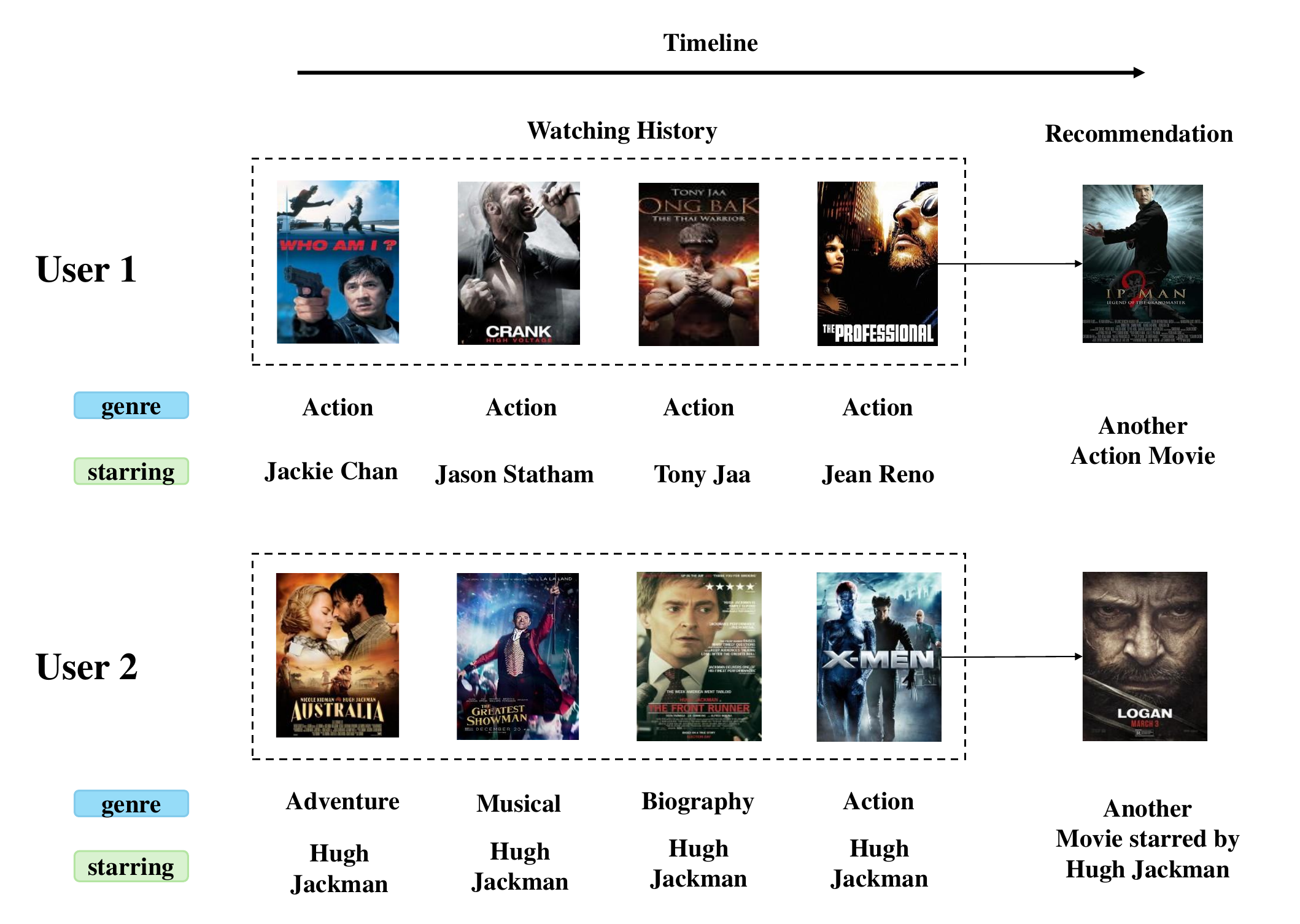}
  \caption{A motivation example.}
  \label{fig: example}
  \Description{A motivation example}
\end{figure}

Mainstream SR systems assign a single embedding vector to each item, capturing its features. Recurrent neural networks (RNNs)~\cite{GRU4Rec, hidasi2015session}, attention-based models~\cite{SASRec, BERT4Rec, FDSA, Autoint}, graph-based models~\cite{SRGNN, SURGE, DCRec, MSGIFSR}, and others combine these item embeddings into a sequence representation vector to capture the user intent. This representation is used to predict the next item (e.g., selecting the item with the highest inner product with the sequence representation vector).
However, a single embedding cannot well capture an item's multifaceted nature (e.g., movie genres and starring)~\cite{Re4, MiasRec}. This is particularly problematic when different facets of an item can influence user intent.
As illustrated in Figure~\ref{fig: example}, User 1's watch history suggests a strong preference for action movies. {\cheng In this case}, recommending another action movie might be appropriate. Conversely, User 2's movie choices range across genres but all feature Hugh Jackman. In this case, recommending another movie starring Hugh Jackman might be more relevant. These examples highlight how user interests can be dominated by a single facet (genre or actor) within a category (movie).
{\cheng Furthermore, in more realistic scenarios,}
users can have multiple preferences within a single facet. For example, a user might enjoy both action and musical movies in the facet of genre. Recognizing and addressing these diverse preferences within a sequence is crucial for generating more effective recommendations that cater to specific user interests~\cite{MiasRec}.
Failing to capture the dominant facet and the specific preferences within each facet can lead to suboptimal recommendations. This highlights the need for recommender systems that can effectively model the dynamic and multi-faceted nature of user interests.

Existing research addresses user intent complexity {\cheng by} using hierarchical windows~\cite{MSGIFSR, atten-mixer} to capture multi-level user intents from recent items, or {\cheng by} utilizing item representations from multiple items in the sequence instead of using the last item's representation only to recommend the next interacted item~\cite{MiasRec}. However, these methods still neglect the multi-faceted nature of items themselves.

To solve the aforementioned issues,
we propose a novel {\cheng structure} called
\textit{\textbf{\underline{F}}acet-\textbf{\underline{A}}ware \textbf{\underline{M}}ulti-Head Mixture-of-\textbf{\underline{E}}xperts Model for Sequential Recommendation} (\textbf{\textit{FAME}}).
We utilize the sub-embeddings from each head in the final multi-head attention layer to independently predict the next item, with each head capturing a distinct facet of the items. A gating mechanism integrates recommendations from each head {\zhang and} dynamically {\zhang determines} their importance.
Furthermore, we introduce a Mixture-of-Experts (MoE) network: this network replaces the query matrix in the self-attention layer, enabling the model to disentangle various user preferences within each facet. Each expert within the MoE focuses on a specific preference {\zhang within} each facet. A learnable router network is adopted to compute the importance weight for each expert and aggregate them. (e.g., whether action or musical movies are the stronger preference in genre). 

To summarize, our contributions in this paper are as follows:
\begin{itemize}
    \item We propose a Multi-Head Prediction Mechanism to enhance the recommendation quality. This design facilitates capturing the potential multi-facet features of the items.
    \item We propose a Mixture-of-Experts (MoE) network that improves user preference modeling by disentangling multiple preferences in each facet within a sequence. This module seamlessly integrates with existing attention-based models.
    \item Our model demonstrates significant effectiveness compared to various baseline categories (sequential, pre-trained, multi-intent) on four public datasets.
\end{itemize}

%% file: relatedwork.tex
\section{Related Work}
\subsection{Sequential Recommendation}
Recent advancements in neural networks and deep learning have spurred the development of various models to extract rich latent semantics from user behavior sequences and generate accurate recommendations. Convolutional Neural Networks (CNNs)~\cite{Caser}, Recurrent Neural Networks (RNNs)~\cite{GRU4Rec}, Transformer-based models~\cite{SASRec, BERT4Rec, FDSA, Autoint}, and Graph Neural Networks (GNNs)~\cite{SRGNN, SURGE, DCRec, MSGIFSR} have been widely employed to enhance representation learning and recommendation performance.
Self-supervised learning (SSL) has emerged as a promising technique for sequential recommendation~\cite{CL4SRec, ICLRec, Re4, DuoRec, DCRec, SSLRec}, with methods like CL4SRec~\cite{CL4SRec} and ICLRec~\cite{ICLRec} employing data augmentation and contrastive learning to improve sequence representations and capture user intents. Additionally, research has focused on modeling multiple user intents, such as the hierarchical window approach in MSGIFSR~\cite{MSGIFSR} and Atten-Mixer~\cite{atten-mixer}, or the multi-item-based representation in MiasRec~\cite{MiasRec}.
Furthermore, incorporating auxiliary information like item categories or attributes~\cite{cai2021category, S3-Rec} and textual descriptions~\cite{rajput2024recommender} has been explored to enrich item representations. The integration of large language models (LLMs) is another emerging trend in the field~\cite{LLM1, LLM2, llamarec}. However, these approaches are beyond the scope of this paper.

\subsection{Sparse Mixtures of Experts (SMoE)}
The Mixture-of-Experts (MoE) architecture has emerged as a powerful tool for handling complex tasks by distributing computations across multiple specialized models, or experts. While MoE models can significantly enhance model capacity, their computational overhead due to routing data to all experts can be prohibitive. To address this, Sparse Mixture of Experts (SMoE) was introduced, enabling each data point to be processed by a carefully selected subset of experts~\cite{MOE1, MOE2, switchtrans}. This approach offers the potential for substantial computational savings without compromising performance.

While SMoE has shown promise in various domains, its application in sequential recommendation remains relatively under-explored. 
Leveraging SMoE in this context could unlock new opportunities to enhance recommendation quality by effectively capturing and modeling diverse user preferences within a sequence.

%% file: Preliminaries.tex
\section{Preliminaries}

\subsection{Notations and Problem Statement}
Let $\mathcal{U}$ and $\mathcal{V}$ represent the user set and item set, where $u\in \mathcal{U}$ (resp. $v\in \mathcal{V}$) denotes an individual user (resp. item). Consequently, $|\mathcal{U}|$ and $|\mathcal{V}|$ denote the {\cheng sizes} of user set and item set, {\cheng respectively}. For each user $u$, we define their interaction sequence $\mathcal{S}_u = \{ v_{1}^{(u)}, \cdots, v_{i}^{(u)}, \cdots, v_{t}^{(u)}\}$ as a chronologically ordered list of items. Here, $v_{i}^{(u)}\in \mathcal{V}$ represents the item that user $u$ interacted with at time step $i$, and $t$ denotes the length of the interaction sequence for user $u$. 
Given a user interaction sequence $\mathcal{S}_u$, the goal of sequential recommendation is to predict the item that user $u$ will interact with at the next time step, $t+1$. Formally, we can define the problem as:
\begin{equation}
    v_{u}^{(*)}=\mathop{\arg\max}\limits_{v_i \in \mathcal{V}} P(v_{t+1}^{(u)}=v_i \mid \mathcal{S}_u)
\end{equation}

\subsection{Multi-Head Self-Attention}
\label{multiheadselfatt}
\begin{enumerate}
    \item \textbf{Item Embeddings:} The model first obtains embeddings for each item in the sequence (denoted as $x\in \mathbb{R}^{d}$).
    \item \textbf{Query, Key, Value Vectors:} Each item embedding ($x$) is then projected into three vectors:
    \begin{itemize}
        \item Query Vector ($q$): Represents what the model is currently looking for in the sequence.
        \item Key Vector ($k$): Captures the content of the current item.
        \item Value Vector ($v$): Contains the actual information associated with the item.
    \end{itemize}
    These projections are calculated using three trainable weight matrices (denoted by $W_Q, W_K\in \mathbb{R}^{d\times d_{k}}, W_V\in \mathbb{R}^{d\times d_{v}}$):
    \begin{equation}
            \label{eq: KQV}
            q=x^T\cdot W_Q,\quad k=x^T\cdot W_K,\quad v=x^T\cdot W_V,
    \end{equation}
    where $d_{k}$ is the dimension of query and key vector, and $d_{v}$ is the dimension of value vector.
    \item \textbf{Attention Scores:} The model calculates an attention score ($\alpha_{ij}$) for each pair of items $(i, j)$ in the sequence. This score reflects the similarity between the current item's query vector ($q_i$) and the key vector ($k_j$) of each other item. A normalization term ($\sqrt{d}$) is used to account for the vector dimension. The attention scores are then normalized using a softmax function (denoted by $\tilde{\alpha_{ij}}$) to create a probability distribution across all items, indicating the relative importance of each item to the current one.
    \begin{equation}
        \alpha_{ij} = \frac{q_{i}^{T}\cdot k_{j}}{\sqrt{d}},\quad \tilde{\alpha_{ij}}=\frac{exp(\alpha_{ij})}{\sum_{j=1}^{t}exp(\alpha_{ij})}
    \end{equation}
    \item \textbf{Item representation:} {\cheng The} item representation $f_i$ is calculated based on weighted sum of value vectors in the sequence. The weights for this summation are derived from the previously calculated attention scores ($\tilde{\alpha_{ij}}$):
    \begin{equation}
        f_i = \sum_{j=1}^{t}\tilde{\alpha_{ij}}\cdot v_j
        \label{eq: f}
    \end{equation}
\end{enumerate}

%% file: Methods.tex
\section{Methods}

\subsection{Overview}
\begin{figure*}[h]
  \centering
  \includegraphics[width=\linewidth]{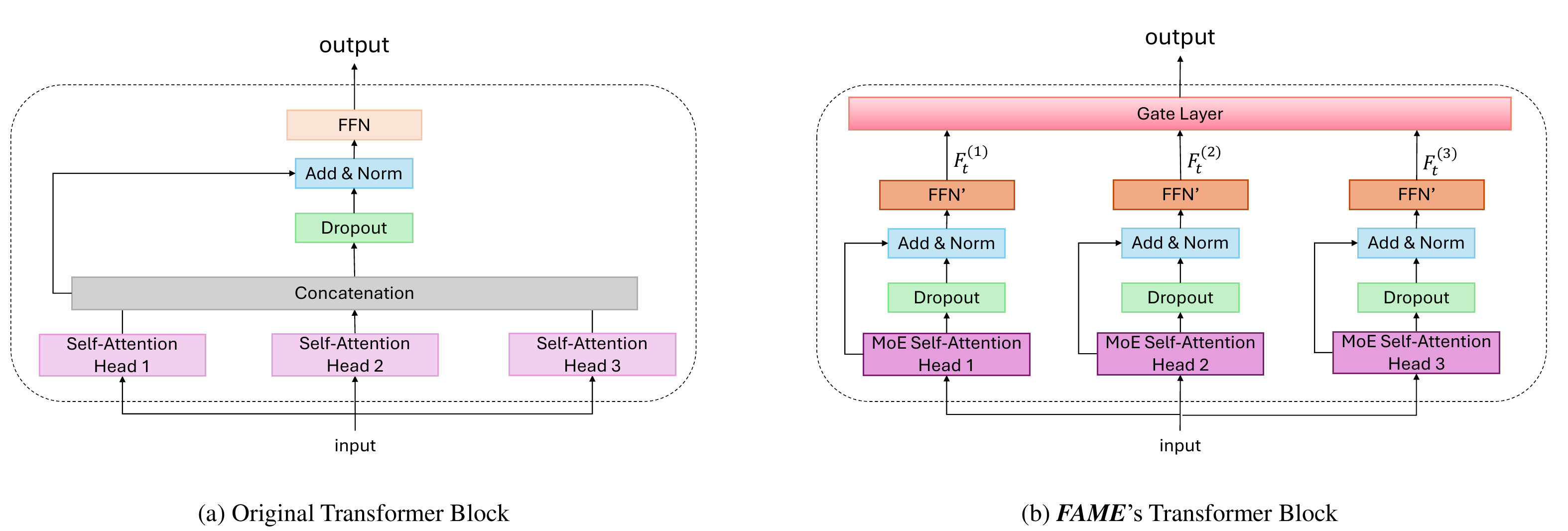}
  \caption{Overview of the proposed model: {\cheng (a)} illustrates the original Transformer block, while {\cheng (b)} depicts the architecture of our proposed \textit{\textbf{FAME}} model. For simplicity, the LayerNorm and Dropout operations following the FFN (FFN') are omitted from the Figure}
  \label{fig: overview}
  \Description{The overview of the model}
\end{figure*}

This section introduces our proposed framework with a high-level overview, which is displayed in Figure~\ref{fig: overview}. 
The framework incorporates two key components: the \textbf{Facet-Aware Multi-Head Prediction Mechanism} (detailed in Section.~\ref{MultiHeadPLayer}), which learns to represent each item with multiple sub-embedding vectors, each capturing a specific facet of the item; and the \textbf{Mixture-of-Experts Self-Attention Layer} (detailed in Section.~\ref{MOEAttLayer}), which employs a Mixture-of-Experts (MoE) network within each subspace to capture the users' specific preferences within each facet.
{\cheng Our framework can be seamlessly integrated to any attention-based recommendation model. In this paper, we incorporate our framework to SASRec for illustration.}

\subsection{Facet-Aware Multi-Head Prediction Mechanism}
\label{MultiHeadPLayer}
\subsubsection{Original SASRec Prediction Process}
\label{SASRec}
In the original SASRec model, the final prediction for the next item is based on the last item's representation ($f_{t}$, calculated by Equation~\ref{eq: f}, which can also be regarded as the sequence representation) obtained from the last self-attention layer. This representation is processed through a feed-forward network (FFN) with ReLU activation for non-linearity, followed by layer normalization, dropout, and a residual connection:
\begin{equation}
    \begin{split}
        {\rm FFN}(f_{t})={\rm RELU}(f_{t}^{T}\cdot W_1 + b_1)^{T}\cdot W_2 + b_2,\\
        F_t={\rm LayerNorm}(f_{t} + {\rm Dropout}({\rm FFN}(f_{t}))),
    \end{split}
    \label{eq: FFN}
\end{equation}
Here, $W_1, W_2\in \mathbb{R}^{d\times d}, b_1, b_2\in \mathbb{R}^{d}$ are all learnable parameters.
The final user preference score for item $v$ at step $(t+1)$ is then calculated as the dot product between the item embedding ($x_v$) and the sequence representation ($F_t$):
\begin{equation}
    P(v_{t+1}=v|\mathcal{S}_u)=x_{v}^{T}\cdot F_t,
\end{equation}
Top-$k$ items with the highest preference scores are recommended to the user.

\subsubsection{Motivation for Our Approach}
\label{multiheadmotivation}

The multi-head self-attention mechanism splits the sequence representation and item embeddings into multiple subspaces (heads). Research suggests that these heads can allocate different attention distributions so as to perform different tasks~\cite{transformer}. We hypothesize that these heads could also capture different facets of items (e.g., genre and starring actors in the context of movie recommendation). This ability to capture multi-faceted information has the potential to improve recommendation quality.

\subsubsection{Proposed Multi-Head Recommendation}
Instead of performing a single attention function with $d$-dimensional keys, values and queries, it is found beneficial to linearly project the queries, keys and values $H$ 
times with different, learned linear projections to $d_{k}$, $d_{k}$ and $d_{v}$ dimensions, respectively~\cite{transformer}.
{\cheng Here, $H$ is the number of heads, and 
$d_{k}$, $d_{k}$ and $d_{v}$ are typically set to $d'=\frac{d}{H}$.}

Leveraging the multi-head attention mechanism, we propose a novel approach where each head independently generates recommendations. The final item embedding from head $h$ is denoted as $f_{t}^{(h)}\in \mathbb{R}^{d'}$. We then process this embedding similarly {\cheng as we do for} the original model:
\begin{equation}
    \begin{gathered}
        {\rm FFN'}(f_{t}^{(h)})={\rm RELU}(f_{t}^{(h)T}\cdot W'_1 + b'_1)^{T}\cdot W'_2 + b'_2,\\
        F_{t}^{(h)}={\rm LayerNorm}(f_{t}^{(h)} + {\rm Dropout}({\rm FFN'}(f_{t}^{(h)}))),
    \end{gathered}
    \label{eq: FFN'}
\end{equation}
Unlike the original FFN (Equation~\ref{eq: FFN}), the feed-forward network applied to each head (${\rm FFN'}$) operates on a reduced dimension of $d'$. The learnable parameters for ${\rm FFN'}$ are therefore adjusted accordingly: $W'_1, W'_2\in \mathbb{R}^{d'\times d'}, b'_1, b'_2\in \mathbb{R}^{d'}$. This adaptation aligns with the dimensionality of sub-embeddings within each head.
To enhance parameter efficiency and improve performance, we adopt a shared feed-forward network (${\rm FFN'}$) across all attention heads.
Each head generates the preference score for each item independently, i.e., 
\begin{equation}
    P^{(h)}(v_{t+1}=v|\mathcal{S}_u)=x_{v}^{(h)T}\cdot F^{(h)}_{t},
\end{equation}
where $x_{v}^{(h)}\in \mathbb{R}^{d'} $ is the sub-embedding of the item $v$, reflecting the features of the specific facet corresponding to the attention head $h$. Specifically, it is calculated by a linear transformation from its original embedding:
\begin{equation}
    x_{v}^{(h)} = x_{v}^{T} \cdot W_{f}^{(h)},
\end{equation}
with $W_{f}^{(h)}\in \mathbb{R}^{d\times d'}$ being a learnable matrix.

In order to integrate the recommendation results from each head, we employ a gate mechanism to determine the relative importance of each head's recommendations:

\begin{equation}
    \begin{gathered}
        g=\left[F_{t}^{(1)}|\dots |F_{t}^{(H)}\right]^T\cdot W_g + b_g, \\
            \tilde{g}=softmax(g)
    \end{gathered}
    \label{eq: gate}
\end{equation}

Here, $\left[\cdot|\cdot\right]$ denotes the concatenation operation. Each element $\tilde{g}^{(h)}\in [0,1]$ within the vector $\tilde{g}$ represents the importance of head $h$ in determining the user's dominant interest or preference. For instance, a higher $\tilde{g}^{(h)}$ for a genre-focused head indicates a stronger preference for specific movie genres, while a higher value for an actor-focused head suggests a preference for movies starring particular actors. The gate mechanism, parameterized by $W_g\in \mathbb{R}^{d\times H}$ and $b_g\in \mathbb{R}^{H}$, learns to assign appropriate weights to each head based on the user's current context.
Finally, we compute a unified preference score for each item by weighting the recommendations from each head:

\begin{equation}
    P(v_{t+1}=v|\mathcal{S}_u)=\sum_{i=1}^{H}\tilde{g}^{(h)}\cdot P^{(h)}(v_{t+1}=v|\mathcal{S}_u)
\end{equation}

This approach allows the model to exploit the strengths of each head while assigning appropriate weights based on their importance in the specific context.

\subsection{Mixture-of-Experts Self-Attention Layer}
\label{MOEAttLayer}
\begin{figure}[htb]
  \centering
  \includegraphics[width=0.85\linewidth]{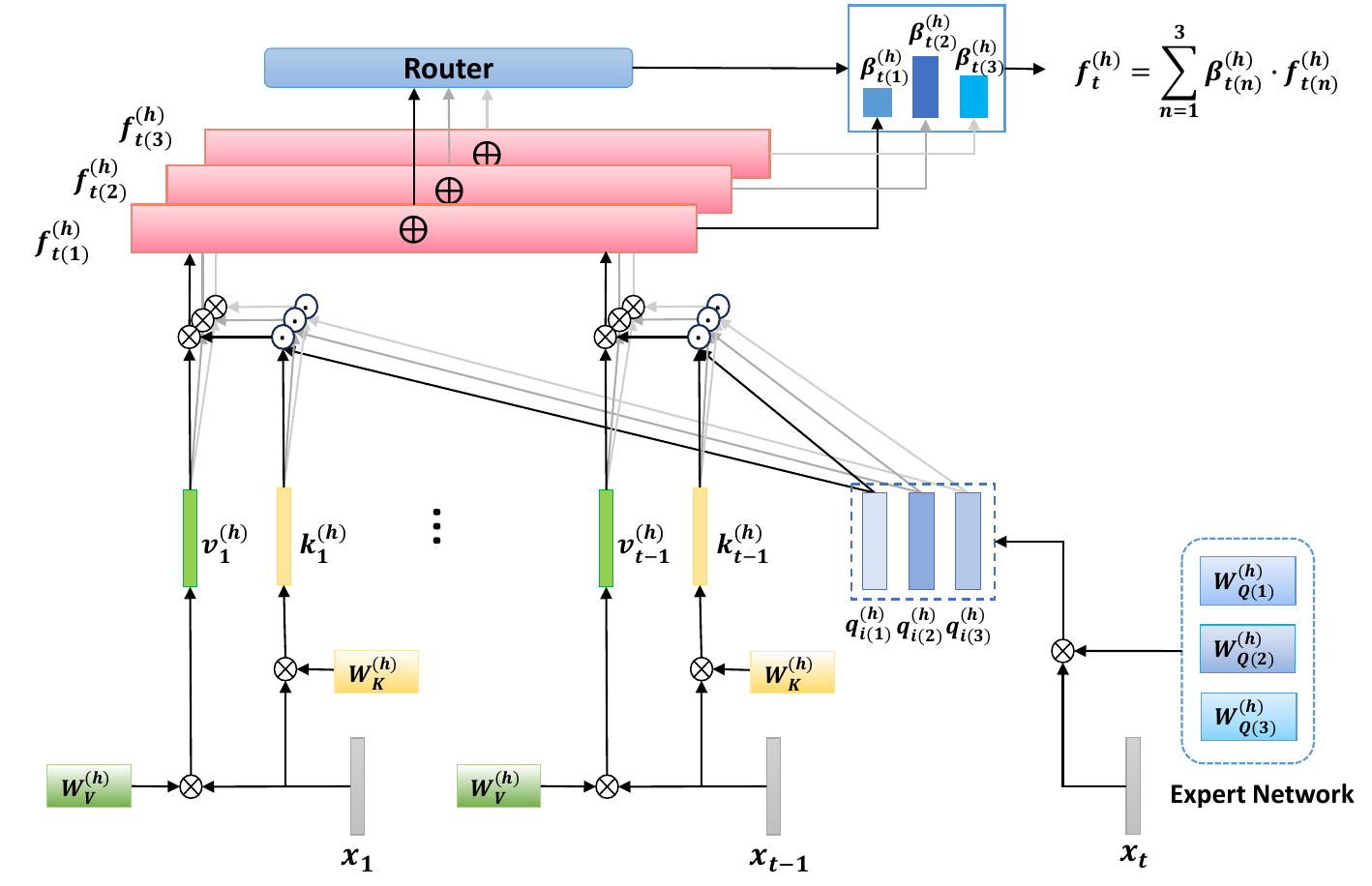}
  \caption{MoE Self-Attention Network: Integrated Item Representation Calculation.
This diagram visualizes the computational process for determining the integrated item representation of the final item ($f_{t}^{(h)}$) within a specific head ($h$) of our proposed model.}
  \label{fig: MOE}
\end{figure}

While the Facet-Aware Multi-Head mechanism effectively captures item facets, users often exhibit more granular and diverse preferences within these facets. To address this, we introduce the Mixture-of-Experts Self-Attention Layer (MoE-Attention), as illustrated in Figure~\ref{fig: MOE}. 

We assume that each facet can be decomposed into $N$ distinct preferences. For instance, a genre facet might include preferences for action, comedy, musicals, etc.
To capture the nuanced preferences within each facet of a sequence, we replace the standard query generation mechanism in self-attention (Equation~\ref{eq: KQV}) with a Mixture-of-Experts (MoE) network in each head. This network consists of $N$ experts, each represented by a trainable matrix $W_{Q(n)}^{(h)}\in \mathbb{R}^{d\times d'}$ (where $n\in [1,N]$). Each expert within a head is designed to capture one of these preferences by transforming an item embedding $x_i$ (i.e., the embedding of the $i^{th}$ item in the sequence) into an expert query vector $q_{i(n)}^{(h)}\in \mathbb{R}^{d'}$ as follows:
\begin{equation}
    q_{i(n)}^{(h)}=x_{i}^{T}\cdot W_{Q(n)}^{(h)}
    \label{eq: newq}
\end{equation}
The key vector ($k_{j}^{(h)}$) and value vector ($v_{j}^{(h)}$) of the $j^{th}$ sequence item in head $h$ are computed using the same linear transformations as in the original SASRec model:
\begin{equation}
    k_{j}^{(h)} = x_{j}^{T}\cdot W_{K}^{(h)},\quad v_{j}^{(h)} = x_{j}^{T}\cdot W_{V}^{(h)}
\end{equation}
Then the attention score for the $i^{th}$ item relative to the $j^{th}$ item in head $h$ by expert $n$ is computed as:
\begin{equation}
    \begin{gathered}
        \alpha_{ij(n)}^{(h)} = \frac{q_{i(n)}^{(h)T}\cdot k_{j}^{(h)T}}{\sqrt{d'}},\\ \tilde{\alpha}_{ij(n)}^{(h)} = softmax(\alpha_{i1(n)}^{(h)},\cdots, \alpha_{it(n)}^{(h)})
    \end{gathered}
\end{equation}

\begin{figure}[htb]
  \centering
  \includegraphics[width=0.92\linewidth]{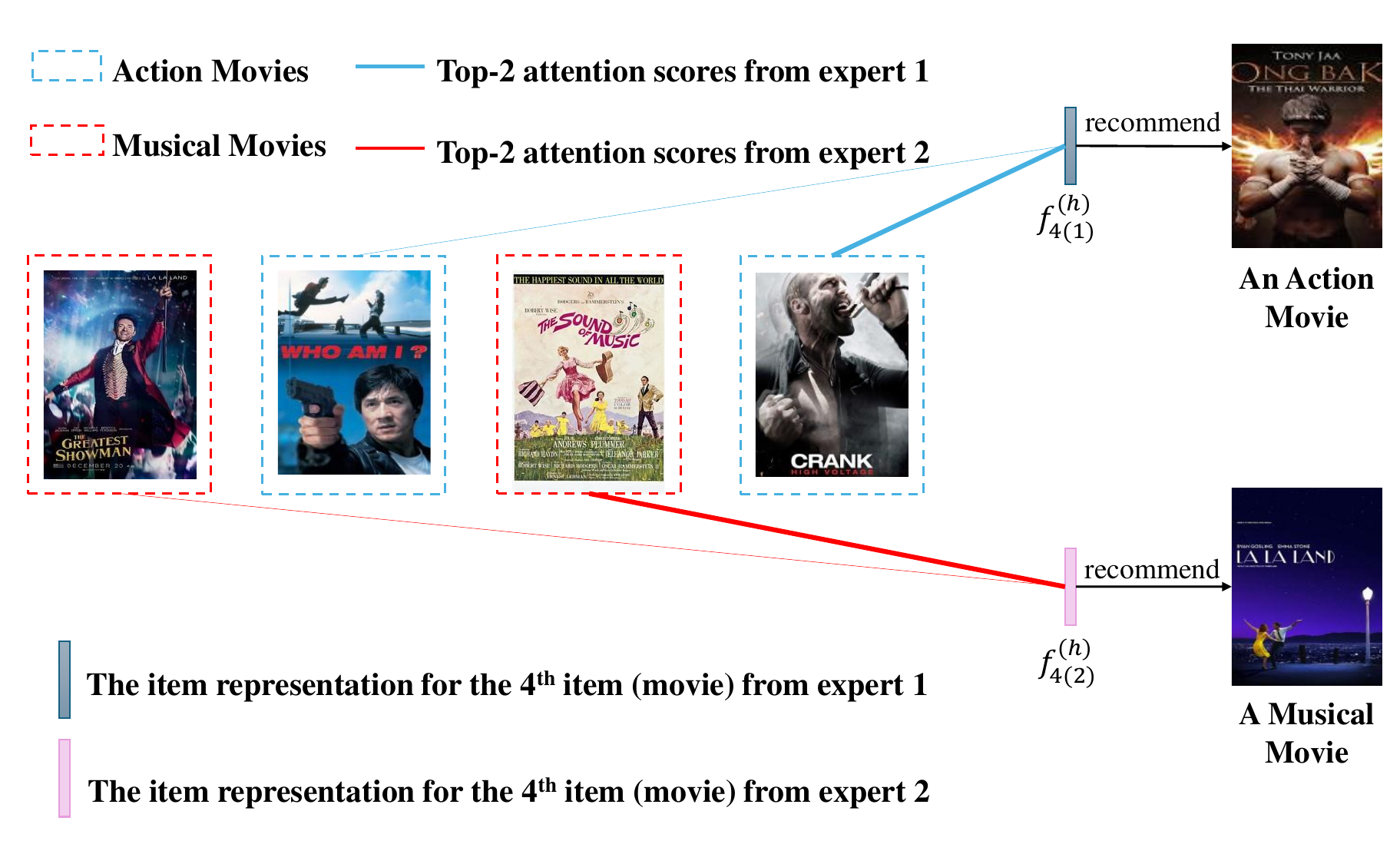}
  \caption{An example on attention scores distribution and recommendation results among different experts on genre-focused head}
  \label{fig: MOEexample}
\end{figure}

The item representation of the $i^{th}$ item for head $h$ and expert $n$ ($f_{i(n)}^{(h)}$) is then calculated as a weighted sum of value vectors, where the weights are the corresponding attention scores:
\begin{equation}
    f_{i(n)}^{(h)} = \sum_{j=1}^{t} \tilde{\alpha}_{ij(n)}^{(h)}\cdot v_{j}^{(h)}
\end{equation}
As illustrated in Figure~\ref{fig: MOEexample}, consider a genre-focused head with two experts: one for action movies and another for musical movies. As detailed in Section.~\ref{SASRec}, the standard SASRec model treats the representation of the final item in a sequence as the overall sequence representation. 
To illustrate our MoE attention mechanism, we focus on the attention scores associated with the $4^{th}$ (and final) item in the sequence. The first expert's query vector of the $4^{th}$ item ($q_{4(1)}^{(h)}$) would assign higher attention scores to action movies (items 1 and 3), while the second expert's query vector ($q_{4(2)}^{(h)}$) would focus on musical movies (items 2 and 4). Consequently, the final item's representation ($f_{4(1)}^{(h)}$) generated by the first expert would lean towards recommending action movies, whereas the representation ($f_{4(2)}^{(h)}$) from the second expert would favor musical movies.

To dynamically determine the importance of each preference within a facet (e.g., whether action or musical is the preferred genre), we introduce a router network parameterized by $W_{exp}^{(h)}\in \mathbb{R}^{(n\cdot d')\times n}$. This network assigns an importance score $\beta_{i(n)}^{(h)}\in (0,1)$ to each item representation generated by each expert $f_{i(n)}^{(h)}$.
The importance scores are computed as follows:
\begin{equation}
    \beta_{i\cdot}^{(h)} = softmax(\left[f_{i(1)}^{(h)}|\cdots |f_{i(n)}^{(h)}\right]^T\cdot W_{exp}^{(h)})
    \label{eq: expertimportance}
\end{equation}
The integrated item representation $f_{i}^{(h)}$ for the $i^{th}$ item in head $h$ is then computed as a weighted sum of the expert query vectors:
\begin{equation}
    f_{i}^{(h)} = \sum_{n=1}^{N}\beta_{i(n)}^{(h)}\cdot f_{i(n)}^{(h)}
    \label{eq: querysum}
\end{equation}
The integrated item representation ($f_{i}^{(h)}$) represents the overall preference at the $i^{th}$ timestamp within head $h$. For instance, for the case in Figure~\ref{fig: MOEexample}, a higher weight for $f_{4(1)}^{(h)}$ (resp. $f_{4(2)}^{(h)}$) would {\cheng push} the model towards recommending action (resp. musical) movies.

\begin{table}[t]
\centering
\caption{Dataset statistic}
\begin{tabular}{cccccc}
\toprule
Dataset    & \#users & \#items & \#actions & avg.length & density \\
\midrule
Beauty     & 22,363  & 12,101  & 198,502   & 8.8        & 0.07\%  \\
Sports     & 25,598  & 18,357  & 296,337   & 8.3        & 0.05\%  \\
Toys     & 19,412  & 19,392  & 167,597   & 8.6        & 0.04\%  \\
ML-20m       & 96,726  & 16,297  & 185,6746   & 19.2        & 0.11\%  \\
\bottomrule
\end{tabular}
\vspace{-1em}
\label{tab: datasetstat}
\end{table}
\subsection{Deployment and Training}
\subsubsection{Model Deployment}
Our FAME model is built upon the SASRec (or any attention-based) framework, with the final Transformer layer replaced by our proposed architecture.
\subsubsection{Training Pipeline}
We initiate our model by pre-training an attention-based sequential recommendation model (e.g., SASRec). Subsequently, we replace Transformer block's query matrix at the final layer with our proposed MoE network ({\cheng Section}~\ref{MOEAttLayer}) while retaining the original key and value matrices. The newly introduced components, including the head-specific ${\rm FFN'}$ (Equation~\ref{eq: FFN'}), gate mechanism (Equation~\ref{eq: gate}), and router (Equation~\ref{eq: expertimportance}), are randomly initialized. The entire model is then fine-tuned end-to-end.

\subsubsection{Training Objectives}
A global cross-entropy loss function is employed to optimize the model during training:
\begin{equation}
    \label{eq: lossce}
    \mathcal{L}_{ce} = -\sum_{u\in \mathcal{U}}\log \left(\frac{exp(x_{t+1}^{(u)T}\cdot f_{t}^{(u)})}{\sum_{i}exp(x_{i}^{T}\cdot f_{t}^{(u)})} \right)
\end{equation}

%% file: experiments.tex
\section{Experiments}

\input{maintable}

\subsection{Datasets}

We conduct experiments on four public datasets. \textit{Beauty}, \textit{Sports} and \textit{Toys} are three subcategories of Amazon review data introduced in~\cite{Amazondata}. 
\textit{ML-20m} is a subset of the MovieLens dataset~\cite{movielens}, containing approximately 20 million ratings from 138,493 users on 27,278 movies. 
Following~\cite{S3-Rec, CL4SRec}, only "5-core" sequences are remained in the 4 datasets, in which all users and items have at least 5 interactions. The statistics of the prepared datasets are summarized in Table ~\ref{tab: datasetstat}.

\subsection{Evaluation Metrics}
We rank the prediction on the whole item set without negative sampling~\cite{fullsort}. Performance is evaluated on a variety of evaluation metrics, including Hit Ratio@$k$ (HR@$k$), and Normalized Discounted Cumulative Gain@$k$ (NDCG@$k$) where $k\in \{5, 10, 20\}$.
Following standard practice in sequential recommendation~\cite{ren2020sequential, SASRec, BERT4Rec, S3-Rec}, we employ a \textit{leave-one-out} evaluation strategy: for each user sequence, the final item serves as the test data, the penultimate item as the validation data, and the remaining items as the training data.

\subsection{Baselines}
We compare our proposed method against a set of baseline models as follows:

\begin{itemize}[left=0pt]
        \item \textbf{GRU4Rec}~\cite{GRU4Rec}: it employs a GRU to encode sequences and incorporates a ranking-based loss.
        \item \textbf{SASRec}~\cite{SASRec}: this method is a pioneering work utilizing self-attention to capture dynamic user interests.
        \item \textbf{BERT4Rec}~\cite{BERT4Rec}: this approach adapts the BERT architecture for sequential recommendation using a cloze task.
        \item \textbf{CORE}~\cite{CORE}: it proposes a representation-consistent encoder based on linear combinations of item embeddings to ensure that sequence representations are in the same space with item embeddings.
        \item \textbf{CL4SRec}~\cite{CL4SRec}: this method combines contrastive learning with a Transformer-based model through data augmentation techniques (i.e., item crop, mask, and reorder).
        \item \textbf{ICLRec}~\cite{ICLRec}: this approach improves sequential recommendation by conducting clustering and contrastive learning on user intentions represented by cluster centroids to enhance recommendation.
        \item \textbf{DuoRec}~\cite{DuoRec}: this research investigates the representation degeneration issue in sequential recommendation and offers solutions based on contrastive learning techniques.
        \item \textbf{MSGIFSR}~\cite{MSGIFSR}: it captures multi-level user intents using a Multi-granularity Intent Heterogeneous Session Graph. 
        \item \textbf{Atten-Mixer}~\cite{atten-mixer}: this method leverages concept-view and instance-view readouts for multi-level intent reasoning instead of using the GNN propagation.
        \item \textbf{MiasRec}~\cite{MiasRec}: this approach utilizes multiple item representations in the sequence instead of only using the last item's representation as the sequence representation to capture diverse user intents.
\end{itemize}

\begin{figure*}[t]
\centering
     \subfloat[Beauty]{\includegraphics[width=0.25\linewidth]{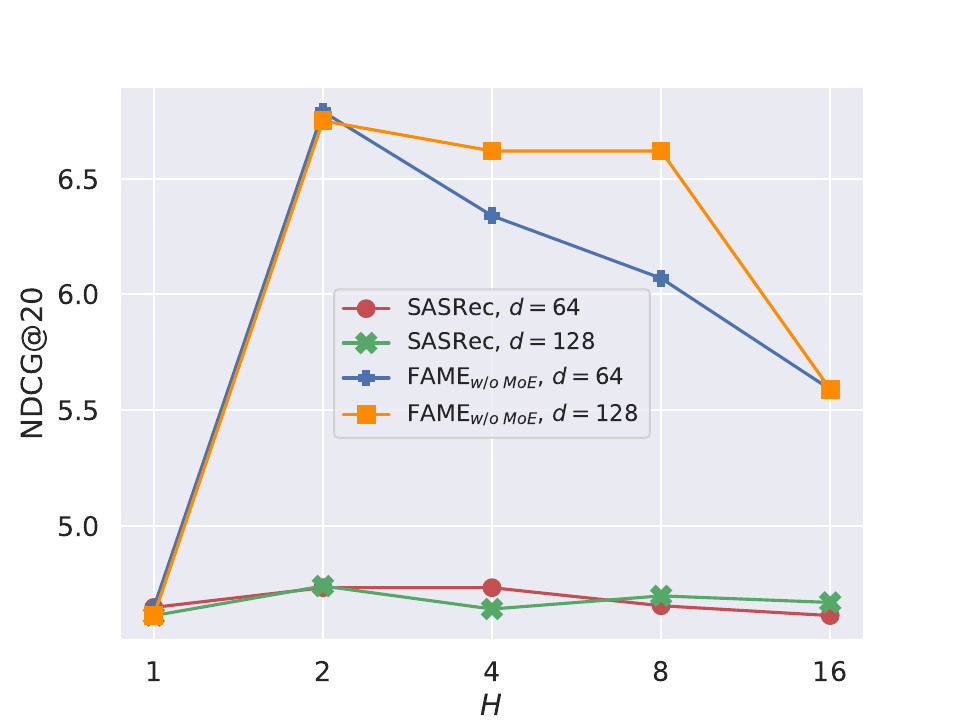}\label{fig: beautyNDCG}}
    \subfloat[Sports]{\includegraphics[width=0.25\linewidth]{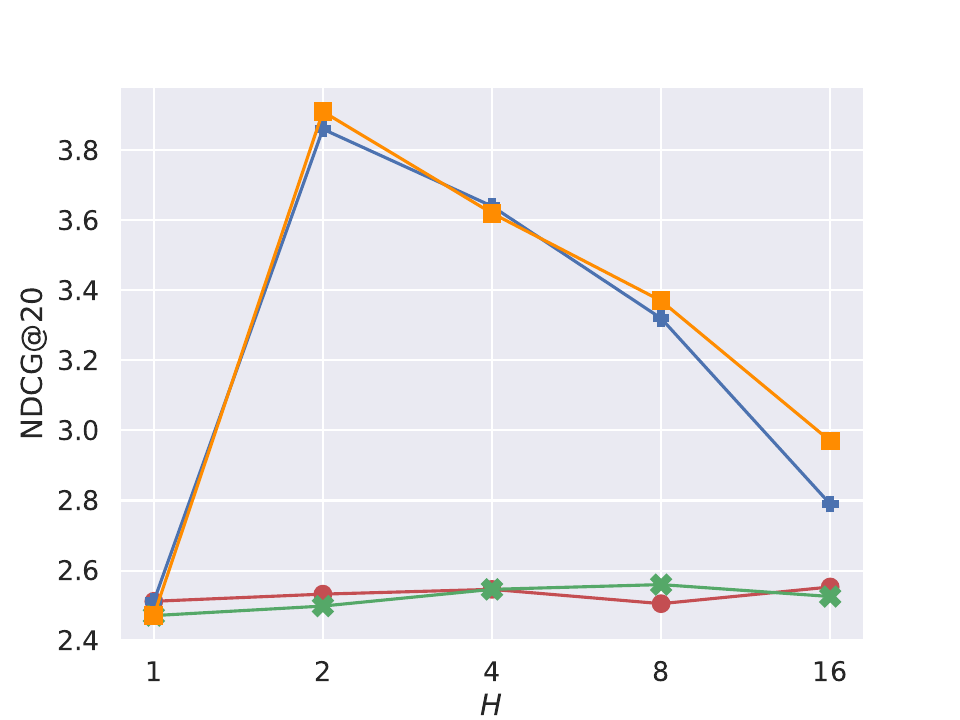}\label{fig: sportsNDCG}}
	\subfloat[Toys]{\includegraphics[width=0.25\linewidth]{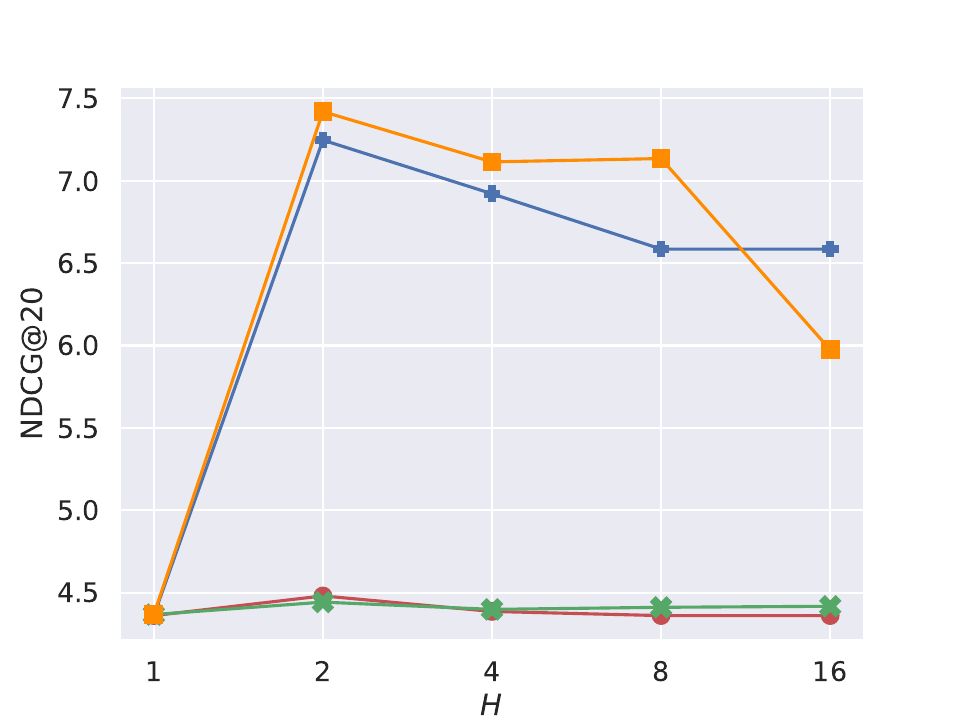}\label{fig: toysNDCG}}
	\subfloat[ML-20m]{\includegraphics[width=0.25\linewidth]{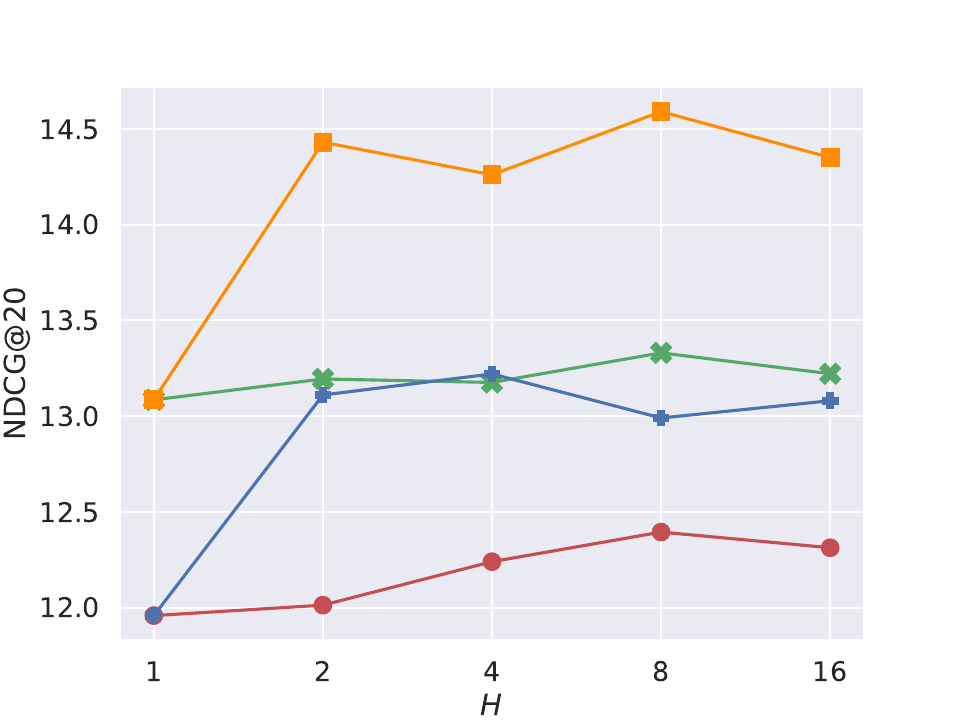}\label{fig: ml20mNDCG}}\\
    \subfloat[Beauty]{\includegraphics[width=0.25\linewidth]{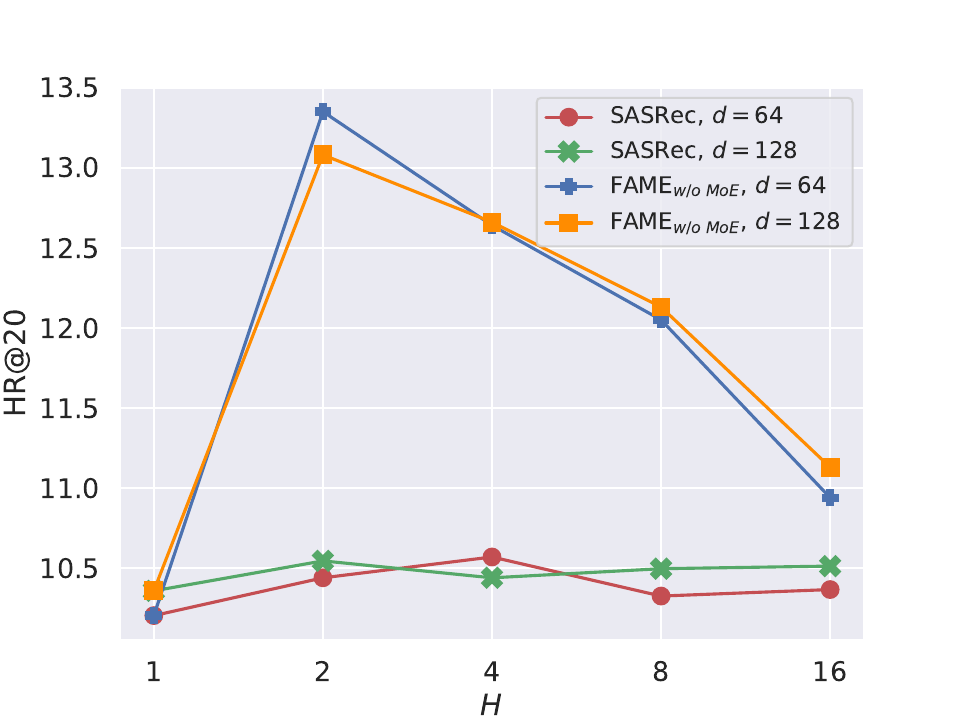}\label{fig: beautyHR}}
    \subfloat[Sports]{\includegraphics[width=0.25\linewidth]{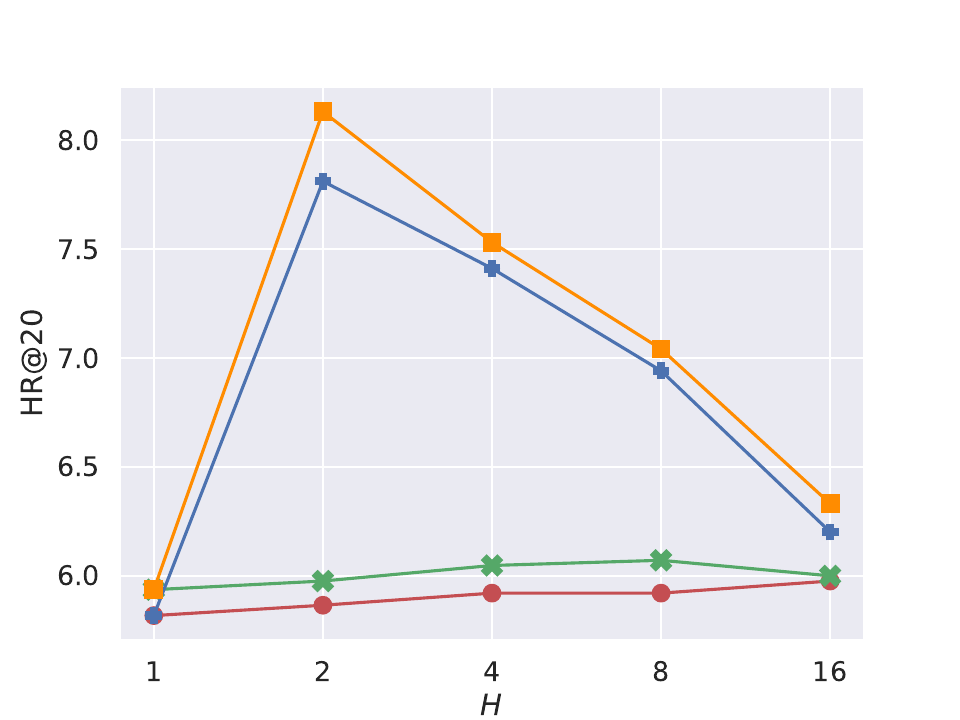}\label{fig: sportsHR}}
	\subfloat[Toys]{\includegraphics[width=0.25\linewidth]{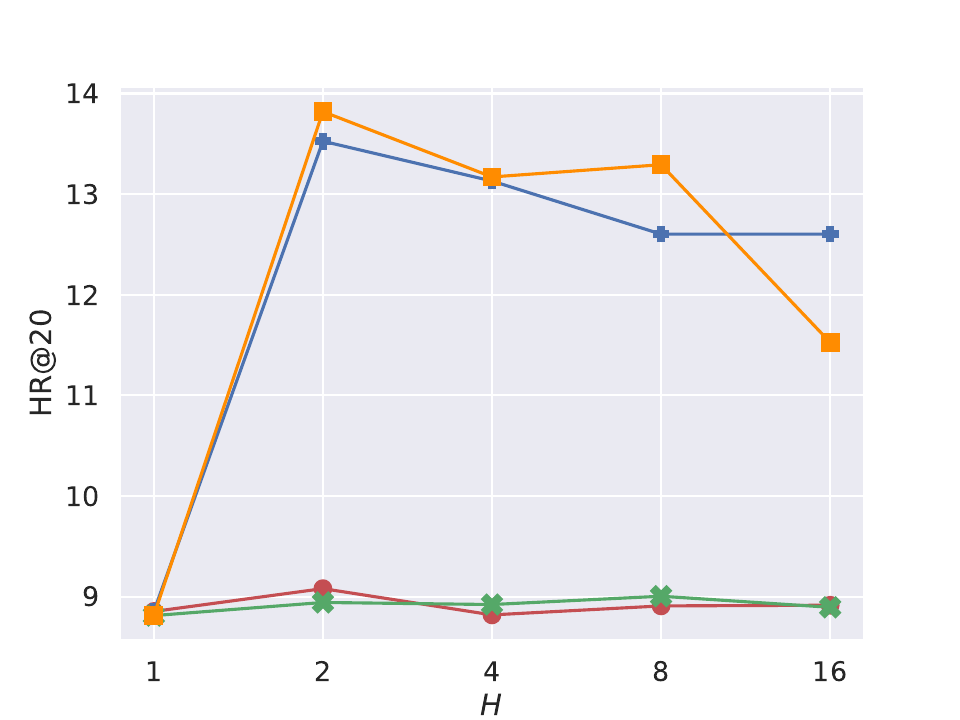}\label{fig: toysHR}}
	\subfloat[ML-20m]{\includegraphics[width=0.25\linewidth]{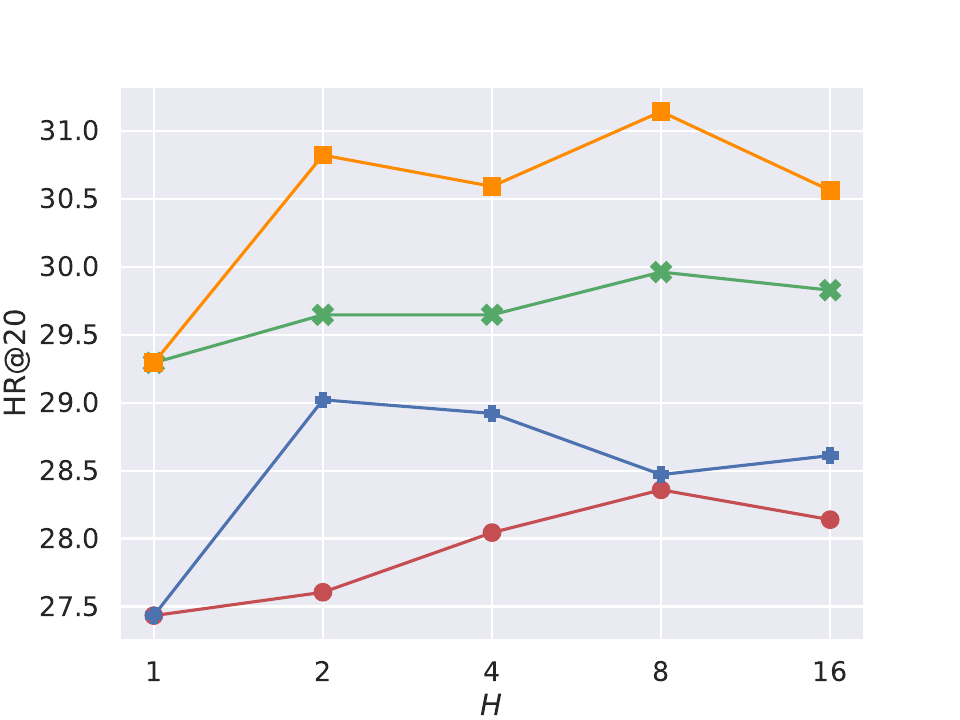}\label{fig: ml20mHR}}
\caption{The performances comparison varying the number of heads in each dataset. The metric in (a)-(d) is NDCG@20, and the metric in (e)-(h) is HR@$20$.}
\label{fig: head study}
\end{figure*}

\begin{figure*}[t]
\centering
     \subfloat[Beauty]{\includegraphics[width=0.25\linewidth]{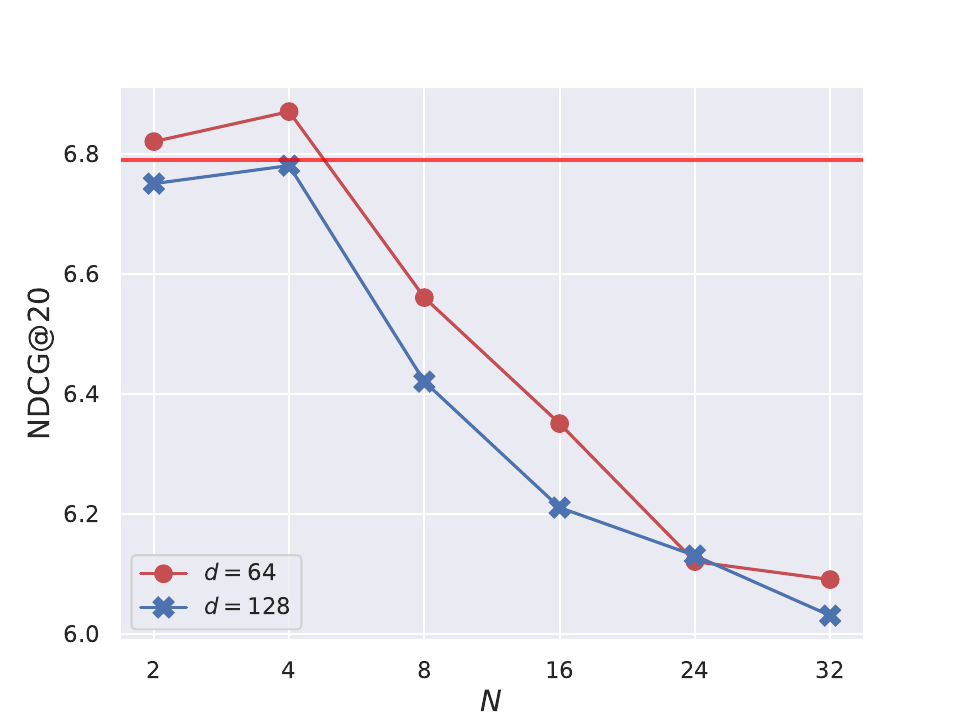}\label{fig: beautyNDCGMoE}}
    \subfloat[Sports]{\includegraphics[width=0.25\linewidth]{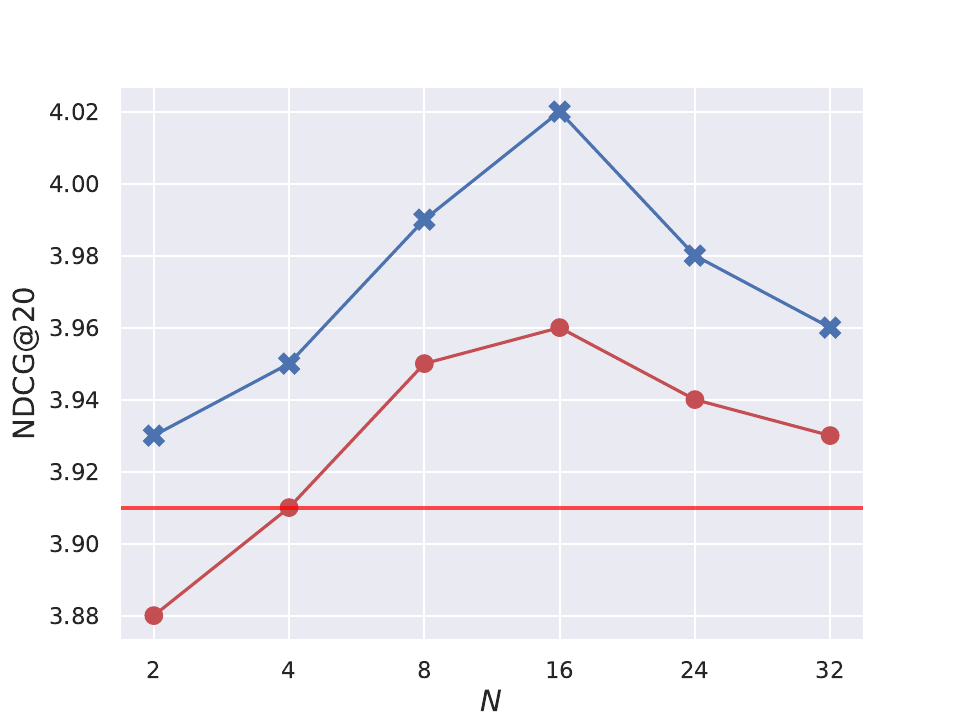}\label{fig: sportsNDCGMoE}}
	\subfloat[Toys]{\includegraphics[width=0.25\linewidth]{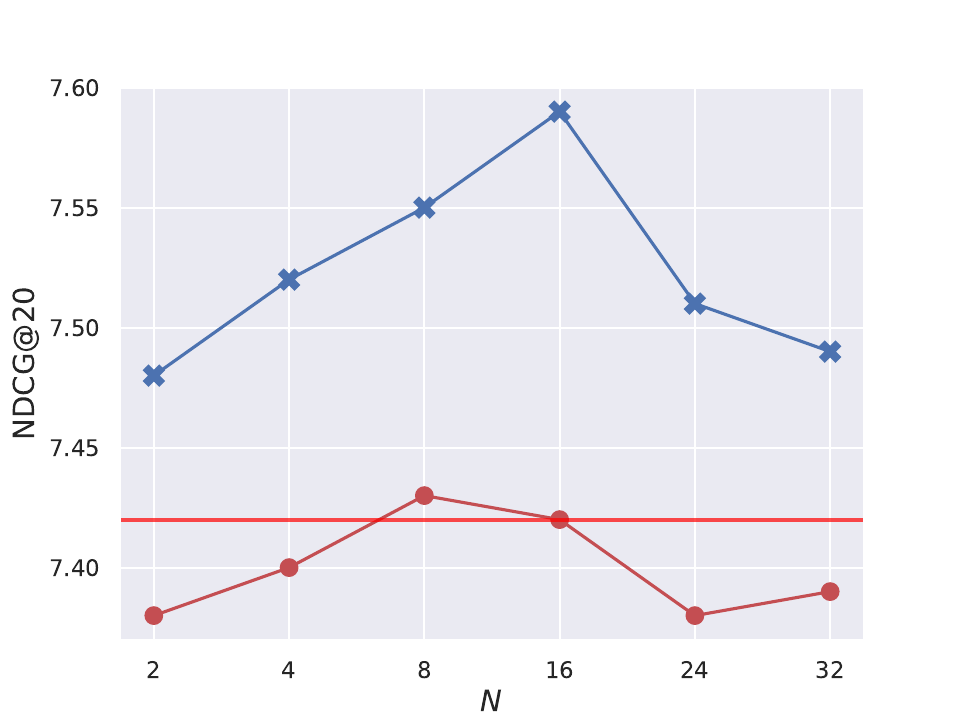}\label{fig: toysNDCGMoE}}
	\subfloat[ML-20m]{\includegraphics[width=0.25\linewidth]{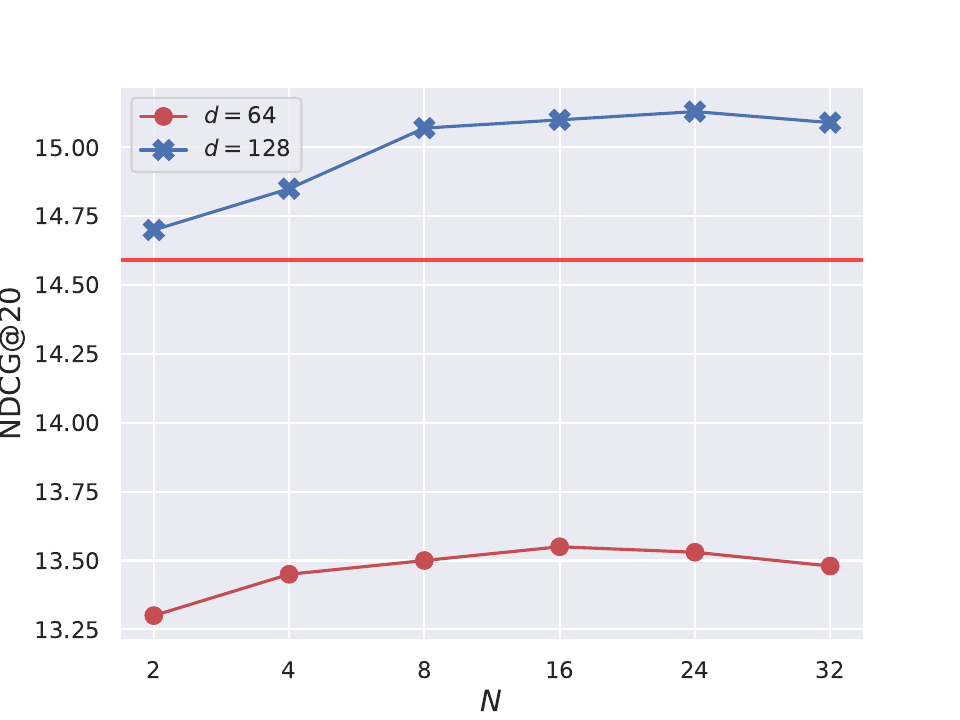}\label{fig: ml20mNDCGMoE}}\\
    \subfloat[Beauty]{\includegraphics[width=0.25\linewidth]{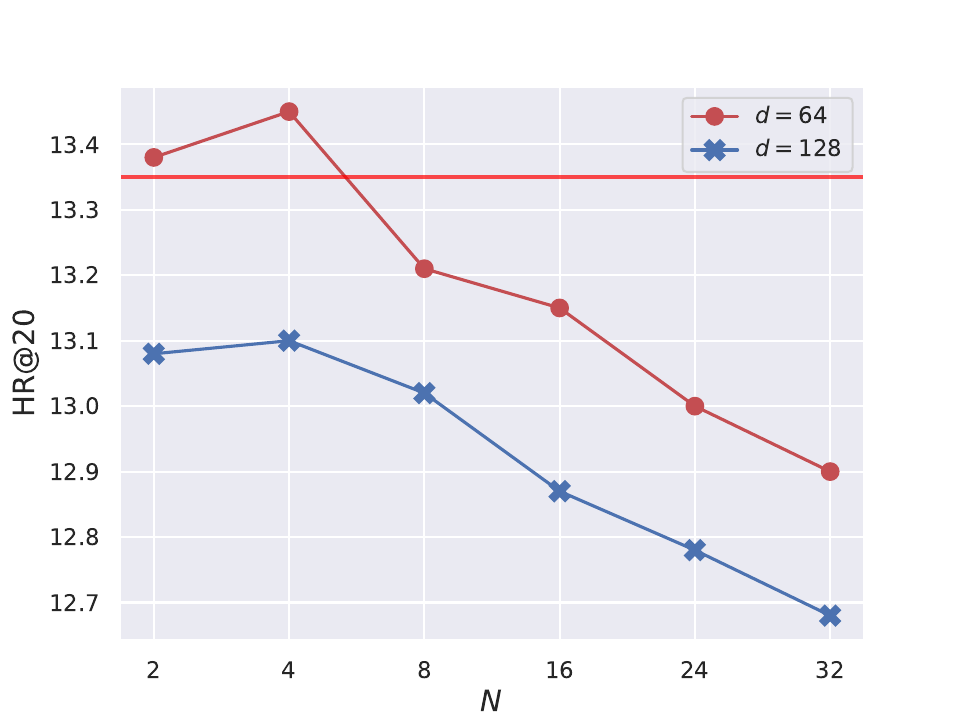}\label{fig: beautyHRMoE}}
    \subfloat[Sports]{\includegraphics[width=0.25\linewidth]{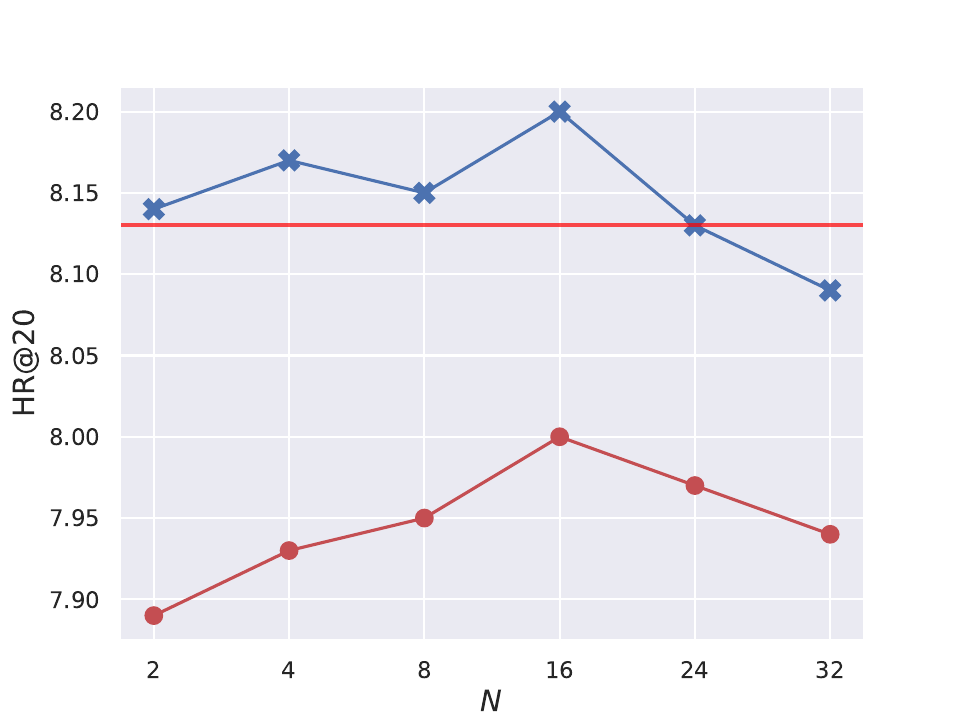}\label{fig: sportsHRMoE}}
	\subfloat[Toys]{\includegraphics[width=0.25\linewidth]{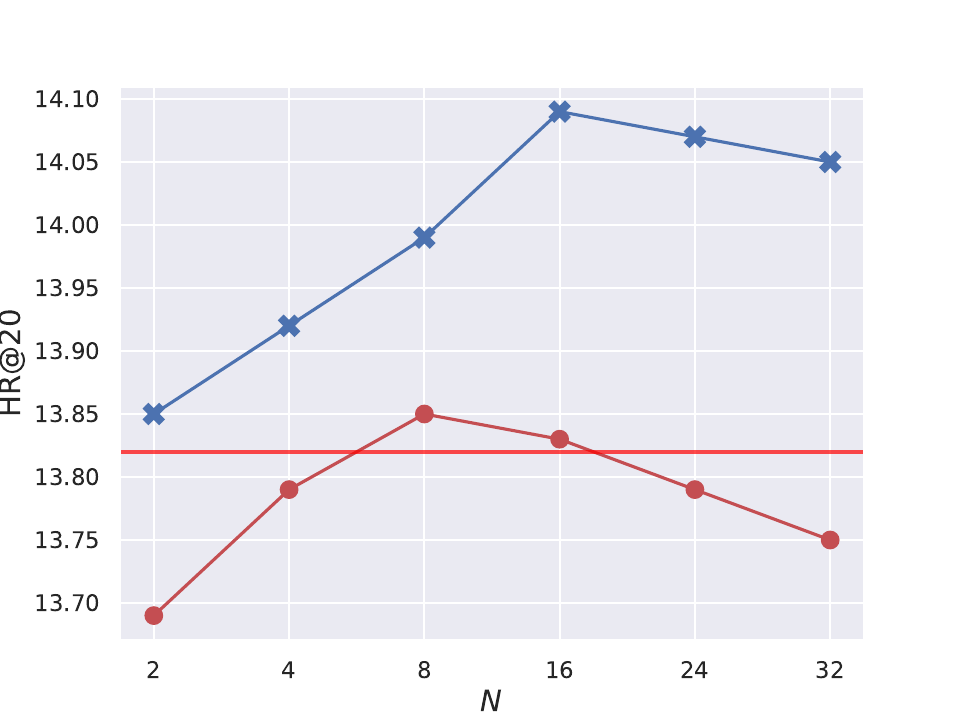}\label{fig: toysHRMoE}}
	\subfloat[ML-20m]{\includegraphics[width=0.25\linewidth]{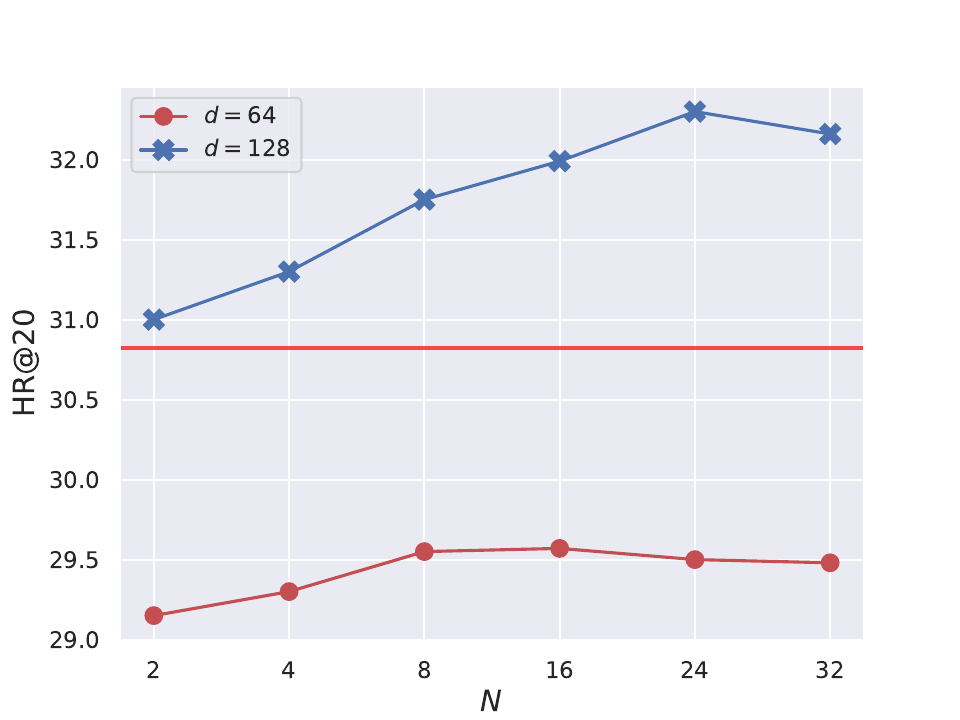}\label{fig: ml20mHRMoE}}
\caption{The performances comparison varying the number of experts in each dataset. The metric in (a)-(d) is NDCG@20, and the metric in (e)-(h) is HR@20. The red horizontal line in each subfigure indicates the peak performance (NDCG@20 or HR@20) achieved by FAME$_{w/o\; MoE}$ within that dataset, as shown in Figure~\ref{fig: head study}.}
\label{fig: expert study}
\end{figure*}

\begin{figure}[t]
\centering
     \includegraphics[width=\linewidth]{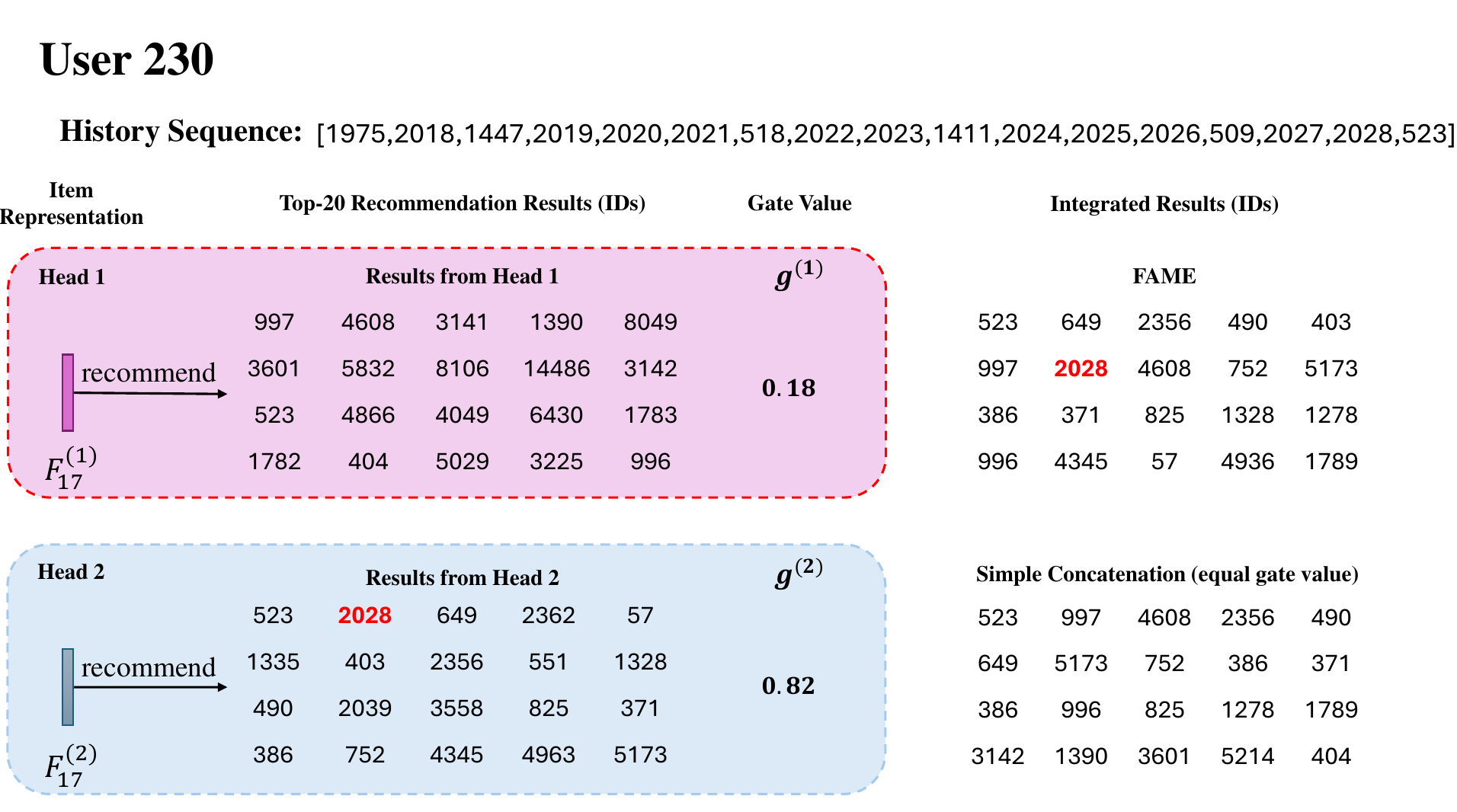}
\caption{Recommendation results for user 230 in the Sports dataset. User history is displayed at the top. The ground truth next item (item 2028) is highlighted.}
\label{fig: case study}
\end{figure}

\subsection{Settings and Implementation Details}
We employ original implementations for SASRec, ICLRec, MSGIFSR, Atten-Mixer, and MiasRec public in their papers. For GRU4Rec and CORE , we leverage the RecBole library\footnote{https://github.com/RUCAIBox/RecBole}~\cite{recbole[1.2.0]}, while BERT4Rec, CL4SRec, and DuoRec are implemented using the SSLRec library\footnote{https://github.com/HKUDS/SSLRec}~\cite{SSLRec}. Hyperparameters for all models are set according to their respective papers.
We experiment with embedding dimensions of 64 and 128 (as experimented, larger dimensions often lead to convergence issues) {\cheng and select} the configuration that yields the best performance for each model.

Our method is implemented in PyTorch. The model is optimized by Adam optimizer with a learning rate of 0.001, $\beta_1 = 0.9, \beta_2 = 0.999$. We employ a batch size of 256. For FAME hyperparameters, $H$ and $N$ are tuned within the ranges $\{1,2,4,8,16 \}$ and $\{2,4,8,16,32 \}$, respectively. All experiments were conducted on a single NVIDIA RTX A5000 GPU.
\subsection{Overall Performance}

Table~\ref{tab: mainresults} presents a comprehensive comparison of FAME against various baseline models. Our experimental findings reveal several key observations:
\begin{itemize}[left=0pt]
    \item \textbf{Limitations of Traditional Models:} While RNN and Transformer-based models have shown success in sequential tasks, their direct application to recommendation often yields suboptimal results due to a lack of consideration for real-world user and item complexities (e.g., GRU4Rec, BERT4Rec).
    \item \textbf{Importance of Intent Modeling:} 
    Models {\cheng that explicitly capture} user intents significantly outperform traditional sequential models. This improvement is attributed to their ability to: 1) handle noisy user behavior by focusing on underlying preferences rather than superficial interactions (e.g., ICLRec), or 2) disentangle multiple co-existing user intents within a sequence (e.g., MiasRec).
    \item \textbf{Superiority of our model FAME:} Our proposed FAME model consistently outperforms all baselines across datasets. This highlights the importance of considering item multi-faceted nature and disentangling user preferences within each facet for effective sequential recommendation.
\end{itemize}

\subsection{Ablation Study and Parameters Study}
This subsection presents the ablation study to evaluate the contributions of our proposed components and conduct corresponding hyperparameter tuning. We begin by examining FAME$_{w/o\; MoE}$, which excludes the MoE module, to assess the impact of the facet-aware multi-head prediction mechanism (introduced in Section.~\ref{MultiHeadPLayer}) and determine the optimal number of attention heads in Section.~\ref{headexp}. Subsequently, using the optimized head configuration, we evaluate the complete FAME model to validate the effectiveness of the MoE module and identify the optimal number of experts in Section.~\ref{expertexp}.

\subsubsection{Impact of the number of heads}
\label{headexp}

Figure~\ref{fig: head study} illustrates the performance variation with different numbers of heads ($H$), treated as a hyperparameter. We compare the original SASRec model with FAME$_{w/o\; MoE}$ to isolate the impact of our multi-head prediction mechanism. 
We experiment with $H$ values of $\{1,2,4,8,16 \}$. When $H=1$, our FAME$_{w/o\; MoE}$ is reduced to the original SASRec model with single head. 
As noted in~\cite{transformer}, computational costs remain constant when varying the number of heads ($H$) while maintaining a fixed embedding dimension ($d$).
\\\textbf{Benefits of multi-head attention:} Both SASRec and FAME$_{w/o\; MoE}$ exhibit performance improvements with multiple heads, however, excessive heads can lead to diminishing returns, aligning with findings in Transformer~\cite{transformer} and SASRec~\cite{SASRec}.
\\\textbf{Superiority of facet-aware architecture:} FAME$_{w/o\; MoE}$ consistently outperforms SASRec, demonstrating the effectiveness of our facet-aware approach.
\\\textbf{Dataset-specific optimal head count:} The optimal number of heads varies across datasets. Beauty, Sports, and Toys benefit from fewer heads, suggesting simpler item facets, while ML-20m requires more heads to capture complex item characteristics.

\subsubsection{Impact of the number of experts}
\label{expertexp}
Figure~\ref{fig: expert study} illustrates the influence of the number of experts ($N$) within each attention head on model performance. We set $H$ to the optimal value determined for FAME$_{w/o\; MoE}$ and compare its performance (red horizontal line in each subfigure) to {\cheng that of} FAME {\cheng by} varying $N$ in $\{2,4,8,16,32 \}$. 
FAME simplifies to FAME$_{w/o\; MoE}$ when $N$ is set to 1.

FAME outperforms FAME$_{w/o\; MoE}$ across all datasets. This improvement can be attributed to the effectiveness of the MoE component, as evidenced by the existence of an optimal $N$ value in each subfigure that surpasses the performance of FAME$_{w/o\; MoE}$.
While the Beauty dataset exhibits diminishing returns for $N$ greater than 4, suggesting simpler user preferences, the other datasets benefit from a larger number of experts. In particular, ML-20m show performance gains with increasing $N$, indicating the presence of more complex and diverse user preferences. However, excessive experts ($N=32$) might lead to overfitting in the Sports and Toys dataset.

\subsection{Case Study}
To illustrate the effectiveness of our facet-aware mechanism, Figure~\ref{fig: case study} presents recommendation results for user 230, along with corresponding head importance scores (calculated using Equation~\ref{eq: gate}). For simplicity, we set the number of heads to two and focus on the Sports dataset. 

The figure clearly demonstrates the diversity of recommendations across different heads, highlighting the ability of our model to capture distinct item facets. The calculated head importance scores reveal that head 2 better aligns with user 230's preferences (0.82 vs 0.18), as evidenced by the inclusion of the ground truth item (item 2028) in its recommendation list. The integrated recommendation, incorporating both heads with appropriate weights, successfully predicts the ground truth item.

In contrast, a traditional approach concatenating sub-embeddings from all heads without considering head importance fails to capture the user's dominant preference, resulting in the omission of the ground truth item in the recommendation list.



%% file: maintable.tex
\begin{table*}[t]
\small
\centering
\caption{Performance comparison of different methods on top-$k$ recommendation}
\label{tab: mainresults}
\begin{tabular}{clcccccccccccc}
\toprule
Dataset                 & Metric  & GRU4Rec & SASRec       & BERT4Rec     & CORE   & CL4SRec & ICLRec       & DuoRec       & A-Mixer & MSGIFSR & MiaSRec      & FAME            & Improv. \\
\midrule
\multirow{6}{*}{Beauty} & HR@5    & 0.0408  & 0.0508       & 0.0510       & 0.0331 & 0.0623 & {\ul 0.0664} & 0.0504       & 0.0507  & 0.0518  & 0.0524       & \textbf{0.0710} & 6.9\%  \\
                        & HR@10   & 0.0623  & 0.0761       & 0.0745       & 0.0664 & 0.0877 & {\ul 0.0918} & 0.0691       & 0.0752  & 0.0771  & 0.0795       & \textbf{0.0978} & 6.2\%  \\
                        & HR@20   & 0.0895  & 0.1057       & 0.1075       & 0.1071 & 0.1195 & {\ul 0.1252} & 0.0912       & 0.1033  & 0.1105  & 0.1125       & \textbf{0.1345} & 7.4\%  \\
                        & NDCG@5  & 0.0273  & 0.0318       & 0.0343       & 0.0164 & 0.0440 & {\ul 0.0480} & 0.0363       & 0.0350  & 0.0344  & 0.0362       & \textbf{0.0508} & 5.8\%  \\
                        & NDCG@10 & 0.0342  & 0.0400       & 0.0419       & 0.0271 & 0.0521 & {\ul 0.0562} & 0.0424       & 0.0421  & 0.0429  & 0.0449       & \textbf{0.0593} & 5.5\%  \\
                        & NDCG@20 & 0.0410  & 0.0474       & 0.0502       & 0.0373 & 0.0601 & {\ul 0.0646} & 0.0479       & 0.0504  & 0.0508  & 0.0532       & \textbf{0.0687} & 6.3\%  \\
\hline
\multirow{6}{*}{Sports} & HR@5    & 0.0210  & 0.0266       & 0.0252       & 0.0150 & 0.0338 & {\ul 0.0384} & 0.0225       & 0.0217  & 0.0268  & 0.0270       & \textbf{0.0400} & 4.2\%  \\
                        & HR@10   & 0.0339  & 0.0412       & 0.0395       & 0.0342 & 0.0498 & {\ul 0.0543} & 0.0327       & 0.0321  & 0.0425  & 0.0435       & \textbf{0.0580} & 6.8\%  \\
                        & HR@20   & 0.0527  & 0.0618       & 0.0607       & 0.0609 & 0.0723 & {\ul 0.0753} & 0.0476       & 0.0469  & 0.0634  & 0.0651       & \textbf{0.0820} & 8.9\%  \\
                        & NDCG@5  & 0.0136  & 0.0158       & 0.0166       & 0.0072 & 0.0235 & {\ul 0.0266} & 0.0161       & 0.0165  & 0.0171  & 0.0180       & \textbf{0.0277} & 4.1\%  \\
                        & NDCG@10 & 0.0178  & 0.0205       & 0.0212       & 0.0134 & 0.0287 & {\ul 0.0317} & 0.0193       & 0.0188  & 0.0221  & 0.0233       & \textbf{0.0337} & 6.3\%  \\
                        & NDCG@20 & 0.0225  & 0.0256       & 0.0265       & 0.0201 & 0.0344 & {\ul 0.0370} & 0.0231       & 0.0235  & 0.0279  & 0.0288       & \textbf{0.0402} & 8.6\%  \\
\hline
\multirow{6}{*}{Toys}   & HR@5    & 0.0369  & 0.0489       & 0.0464       & 0.0338 & 0.0658 & {\ul 0.0792} & 0.0481       & 0.0565  & 0.0576  & 0.0581       & \textbf{0.0820} & 3.5\%  \\
                        & HR@10   & 0.0524  & 0.0676       & 0.0677       & 0.0699 & 0.0912 & {\ul 0.1043} & 0.0666       & 0.0819  & 0.0831  & 0.0828       & \textbf{0.1065} & 2.1\%  \\
                        & HR@20   & 0.076   & 0.0908       & 0.0968       & 0.1114 & 0.1209 & {\ul 0.1382} & 0.0879       & 0.1099  & 0.1150  & 0.1143       & \textbf{0.1409} & 1.9\%  \\
                        & NDCG@5  & 0.0247  & 0.0329       & 0.0322       & 0.0158 & 0.047  & {\ul 0.0579} & 0.0356       & 0.403   & 0.0407  & 0.0408       & \textbf{0.0603} & 4.1\%  \\
                        & NDCG@10 & 0.0296  & 0.0389       & 0.0391       & 0.0274 & 0.0552 & {\ul 0.0660} & 0.0415       & 0.481   & 0.0492  & 0.0488       & \textbf{0.0681} & 3.2\%  \\
                        & NDCG@20 & 0.0356  & 0.0448       & 0.0464       & 0.0378 & 0.0627 & {\ul 0.0745} & 0.0469       & 0.574   & 0.0577  & 0.0567       & \textbf{0.0759} & 1.9\%  \\
\hline
\multirow{6}{*}{ML-20m} & HR@5    & 0.1365  & 0.1305       & 0.1446       & 0.0655 & 0.1205 & 0.1380       & {\ul 0.1458} & 0.1325  & 0.1303  & 0.1367       & \textbf{0.1538} & 5.5\%  \\
                        & HR@10   & 0.2052  & 0.2016       & {\ul 0.2172} & 0.1312 & 0.1853 & 0.2070       & 0.2164       & 0.2022  & 0.2013  & 0.2071       & \textbf{0.2246} & 3.4\%  \\
                        & HR@20   & 0.2981  & 0.2996       & {\ul 0.3132} & 0.2251 & 0.2760 & 0.2997       & 0.3108       & 0.2994  & 0.2978  & 0.3021       & \textbf{0.3230} & 3.1\%  \\
                        & NDCG@5  & 0.0927  & 0.0858       & 0.0964       & 0.0347 & 0.0804 & 0.0927       & {\ul 0.0986} & 0.0899  & 0.0844  & 0.0918       & \textbf{0.1046} & 6.1\%  \\
                        & NDCG@10 & 0.1148  & 0.1086       & 0.1197       & 0.0558 & 0.1012 & 0.1149       & {\ul 0.1212} & 0.1121  & 0.1119  & 0.1144       & \textbf{0.1276} & 5.3\%  \\
                        & NDCG@20 & 0.1382  & 0.1333       & 0.1438       & 0.0794 & 0.1240 & 0.1382       & {\ul 0.1450} & 0.1334  & 0.1357  & 0.1383       & \textbf{0.1513} & 4.3\% \\
\bottomrule
\end{tabular}
\end{table*}

%% file: conclusion.tex
\section{Conclusion}
In this paper, we propose a \textit{\textbf{\underline{F}}acet-\textbf{\underline{A}}ware \textbf{\underline{M}}ulti-Head Mixture-of-\textbf{\underline{E}}xperts Model for Sequential Recommendation} (\textbf{\textit{FAME}}), leveraging sub-embeddings from each head in the last multi-head attention layer to predict the next item separately. This approach captures the potential multi-faceted nature of items without increasing model complexity. 
A Mixture-of-Experts (MoE) network is adopted in each attention head to disentangle various user preferences within each facet. Each expert within the MoE focuses on a specific preference, and the importance score is calculated by a router network, which is used to aggregate the overall preference. 
Extensive experiments demonstrate the effectiveness of our method over existing baseline models. 